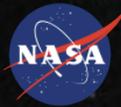

Jet Propulsion Laboratory
California Institute of Technology

Exoplanet Exploration Program
NASA ExEP Mission Star List for the Habitable Worlds Observatory
2023

Eric Mamajek, Deputy Program Chief Scientist
Karl Stapelfeldt, Program Chief Scientist

# NASA ExEP Mission Star List for the Habitable Worlds Observatory: Most Accessible Targets to Survey for Potentially Habitable Exoplanets

Eric Mamajek and Karl Stapelfeldt
NASA Exoplanet Exploration Program (ExEP)
Jet Propulsion Laboratory, California Institute of Technology
January 15, 2023

**Abstract:** The Astro 2020 Decadal Survey "*Pathways to Discovery in Astronomy and Astrophysics for the 2020s*" has recommended that "*after a successful mission and technology maturation program, NASA should embark on a program to realize a mission to search for biosignatures from a robust number of about ~25 habitable zone planets and to be a transformative facility for general astrophysics,*" and prescribing that the high-contrast direct imaging mission would have "*a target off-axis inscribed diameter of approximately 6 meters*." The Decadal Survey assumed an exo-Earth frequency of ~25%, requiring that approximately 100 cumulative habitable zones of nearby stars should be surveyed. Surveying the nearby bright stars, and taking into account inputs from the LUVOIR and HabEx mission studies (but without being overly prescriptive in the required starlight suppression technology or requirements), we compile a list of ~160 stars whose exo-Earths would be the most accessible for a systematic imaging survey of habitable zones with a 6-m-class space telescope in terms of angular separation, planet brightness in reflected light, and planet-star brightness ratio. We compile this star list to motivate observations and analysis to help inform observatory design (mission-enabling "precursor science") and enhance the science return of the *Habitable Worlds Observatory* (HWO) survey for exo-Earths (mission-enhancing "preparatory science"). It is anticipated that this list of target stars and their properties will be updated periodically by the NASA Exoplanet Exploration Program.

**Table of Contents:**

1.  **Motivation**

The Astro 2020 Decadal Survey has recently recommended that "*after a successful mission and technology maturation program, NASA should embark on a program to realize a mission to search for biosignatures from a robust number of about ~25 habitable zone planets and to be a transformative facility for general astrophysics. If mission and technology maturation are successful, as determined by an independent review, implementation should start in the latter part of the decade, with a target launch in the first half of the 2040s.*" (Astro 2020 p.7-17) Astro 2020 concluded that "*...a high-contrast direct imaging mission with a target off-axis inscribed diameter of approximately 6 meters provides an appropriate balance between scale and feasibility. Such a mission will provide a robust sample of ~25 atmospheric spectra of potentially habitable exoplanets, will be a transformative observatory for general astrophysics, and given optimal budget profiles it could launch by the first half of the 2040 decade.*"

There are only a limited number of target stars that can be surveyed by the planned HWO. That number is constrained by the need to achieve the scientific goals articulated above, by a limited number of nearby stars whose habitable zones will be accessible to a plausible 6-m-class space telescope with starlight suppression technologies (e.g., coronagraph, starshade), and stars whose temperate small planets will be bright enough for the telescope to be able to image and spectrally characterize on a reasonable timescale. The Decadal survey further stated that "*[a] space telescope similar in wavelength coverage to the Hubble Space Telescope, and with an aperture of at least 6m and coronagraphic imaging capability should be capable of observing approximately 100 nearby stars, and successfully detect potentially habitable planets around at least a quarter of the systems*." This estimate assumed an "occurrence rate of rocky planets in the optimistic habitable zone" to be $\eta_\oplus$ = 0.24 (Astro 2020 Fig. 7.6, p.7-16, and as adopted in HabEx & LUVOIR study reports). Those corresponding adopted habitable zone limits are a semimajor axis of 0.95-1.67 au for a solar twin, planet sizes between 0.8-1.4 Earth radii, and for non-solar stars, scale as square root of the bolometric luminosity normalized to the Sun. This range of orbital separations correspond to classical limits for the critical fluxes for rocky planets orbiting a G2V star with surface water for cases of water loss and maximum greenhouse (Kasting et al. 1993, Kopparapu et al. 2013), and is dubbed the "*conservative habitable zone.*"

As the proposed Decadal space telescope is intermediate in aperture between the LUVOIR-B (diameter 8m, inscribed diameter 6.7m) and HabEx (4m) concepts, there has not yet been a thorough post-Decadal exploration of the architecture options and impacts on science yields in this intermediate aperture range around 6m. At present, we are informed by the LUVOIR B[1] and HabEx[2] concept studies (2019) which present thorough science cases and instrument concepts.

**The goal of this document and associated data table is to provide the community with a provisional sample of stars which are most likely (given current knowledge) to constitute target stars for the exo-Earth survey of the future Habitable Worlds Observatory mission.** It is hoped that by making this list publicly available that it will motivate community observations and analysis of these nearby stars, which can improve our knowledge of these stars and their companions, improving the fidelity of exoplanet science yield simulations to inform observatory design trade studies and reduce mission risk ("precursor science"), and eventually to inform the final target list and knowledge of these systems in anticipation of the mission itself ("preparatory science").

This list is considerably larger than 100 as the quality of the target stars (in terms of expected exoplanet brightness, angular separation from host star, and other astrophysical considerations like stellar multiplicity, existence of disks, etc.) drops off considerably after the first ~50 stars selected (Tier A), and further analysis of the 2nd and 3rd (Tiers B and C) tier of targets will likely be needed in order to identify the best remaining stars to round out the ~100+ survey targets needed to fulfill a survey for exo-Earths. Yield simulations taking into account a wider range of observatory architecture options, (e.g., aperture), or considering lower values of eta-Earth ($\eta_\oplus$), would benefit from a larger sample than we present here. Stars in such a larger sample may have sufficiently useful photometry, luminosity and distance estimates, but may not yet have undergone the degree of scrutiny that the stars in this target list. The particular stars most amenable to exo-Earth imaging and spectroscopy from yield calculations (e.g., options used in the HabEx & LUVOIR studies) tend to agree between different versions of input catalogs at the ~10% level (C. Stark, priv. comm.), sufficient for yield calculations where uncertainties on parameters like e.g., $\eta_\oplus$ and the Poisson/counting uncertainties, are larger.

---

[1] https://asd.gsfc.nasa.gov/luvoir/ and LUVOIR Team et al. (2019, https://arxiv.org/abs/1912.06219)

[2] https://www.jpl.nasa.gov/habex/documents/ and Gaudi et al. (2020, https://arxiv.org/abs/2001.06683)

## 2. Requirements for Selecting High Priority Stars: Factors and Assumptions

Before describing the assembly and refining of the target star list, we describe several assumptions made about calculating or estimating the properties of the stars, since those properties were used to select and/or remove stars from the list. Several physical or observational stellar parameters were calculated based on other stellar measurements, and used to down-select the candidate target stars.

### Adopted Parameters, Data, Assumptions

- **Passband:** Calculations for estimates of a planet's brightness and the planet-star brightness ratio were done in the Cousins Rc band. Cousins Rc band has $\lambda_{\text{eff}}$ = 642.6 nm for Vega and its throughput for >10% transmission ranges between 554-806 nm (Bessell, Castelli & Plez 1998). Hence this band compares well to the range of ~500-950 nm for LUVOIR's Signature Science Case #1 (Finding habitable planet candidates), between the near-UV/blue-orange visible range (~350-600 nm) which could show Rayleigh scattering or absorption by hazes, and the red/near-IR region (~600-1000 nm) where Earth-like planets might show absorption due to the oxygen A-band (~760 nm) or water vapor (~940 nm). Cousins Rc band also spans the blue (450-670 nm) and red (670-1000 nm) Vis channels for the HabEx coronagraph and Vis 450-1000 nm channel for starshade observations, and between the Habex Objective 1 requirements (~450-550 nm) for detection and orbit determination of Earth-sized planets and HabEx Objective 2 for detection of water vapor (~700-1000 nm). Selecting the stars based on their Rc magnitudes enables a greater sample of later-type targets than a selection based on V band magnitudes.

    While published measurements of Rc photometry were available for many stars (e.g., Cousins 1980, Bessell 1990), for others we adopted published synthetic Rc magnitude estimates from fits of spectral templates to photometry in other bands (e.g., Pickles et al. 2010), or we estimated Rc magnitudes based on Gaia EDR3 G and Rp photometry using relations from Riello et al. (2020). In rare cases, the Rc magnitude was estimated using the Johnson V magnitude and adopting an intrinsic V-Rc color for the star's spectral type following Pecaut & Mamajek (2013). For HIP 64408 and HIP 73996, a Johnson R magnitude from Johnson et al. (1966) was converted to Cousins R (Rc).

    **This work:** Cousins Rc magnitudes were either adopted from the literature or estimated.

- **Geometric Albedo:** Our calculations of the expected brightness of a temperate rocky exoplanet requires an assumed *geometric albedo (p)*, which when combined with a phase function and the physical sizes of the planet and orbital separation, can be used to calculate the brightness ratio of the planet with respect to its star (see section on calculated quantities for planet-star flux ratio). The geometric albedo *p* corresponds to the reflectance of the planet at full phase ($\alpha = 0$) relative to a perfect Lambert disk of the same radius with the same incident flux, where the phase angle $\alpha$ corresponds to the observer-planet-star angle (e.g., Traub & Oppenheimer 2010).

  Both the HabEx and LUVOIR reports (Appendix B.2.5.1) made a similar assumption for the adopted albedo of exo-Earths, and had identical language: "*All exo-Earth candidates were assigned Earth's geometric albedo of 0.2, assumed to be valid at all wavelengths of interest*." The geometric albedo assumed for exoEarth candidates was discussed in Stark et al. (2014) and a value of 0.2 was adopted. For reference, the rocky bodies in the inner solar system have V-band geometric albedos of 0.106 (Mercury), 0.65 (Venus), 0.367 (Earth), 0.12 (Moon), 0.15 (Mars), with the Earth value coming from the review by Harris (1961). Mallama et al. (2017) estimates geometric albedos for Earth in the Johnson-Cousins *VRcIc* bands of 0.434, 0.392, and 0.396, respectively - i.e., all around ~0.4. A campaign of hundreds of observations through 1958 and 1959 by Bakos (1964) yielded visual geometric albedos for Earth of 0.41 (1958) and 0.42 (1959). The classic study by Danjon (1936) measured an albedo of 0.39. Lower albedos are expected for habitable worlds orbiting cooler stars as Rayleigh scattering becomes less important as the starlight spectral energy distributions become increasingly red. The albedo varies very little across the visible for broadband filters, and we assume that the Rc-band albedo and V-band albedos are similar.

  **This work:** Following the HabEx & LUVOIR reports, we adopt a geometric albedo of 0.2 for our calculations in *Rc* band.

- **Earth-Equivalent Instellation Distance (EEID) and Angular Separation:** The orbital separation that receives equivalent irradiance (energy/unit/area) as Earth at 1 au is

  $$r_{EEID} = 1 \text{ au } (L_*/L_{Sun})^{1/2}$$

  (e.g., Stark et al. 2014) where $L_*$ is the host star's bolometric luminosity and $L_{Sun}$ is the IAU nominal solar bolometric luminosity (3.828e26 W). The EEID can also be expressed as an angular separation (quoted in milliarcseconds) corresponding to the maximum angular separation of the planet from the star (for assumed zero-eccentricity circular orbit):

$$\theta_{EEID} = r_{EEID}/d_{pc} = r_{EEID}\, \varpi_{mas}$$

Where $d_{pc}$ is the distance in parsecs, or alternatively calculated using the parallax $\varpi_{mas}$ as $d_{pc} = 1000/\varpi_{mas}$.

- **Habitable Zone (HZ) Limits**: For its calculations of expected Earth-sized planets in the HZ versus effective number of HZs surveyed, the Astro2020 Decadal Survey (Fig. 7.6) adopted a definition of HZ corresponding to orbital radii[3] of 0.95-1.67 AU and planet radii of 0.8-1.4 Earth radii ($R_E$). This follows well the assumptions made in the HabEx and LUVOIR study reports. HabEx: "...*exo-Earth candidates are on circular orbits and reside within the conservative habitable zone (HZ), spanning 0.95–1.67 AU for a solar twin* (Kopparapu et al. 2013)." LUVOIR: "*We adopt the conservative HZ, spanning 0.95–1.67 AU for a solar-twin star (Kopparapu et al. 2013)*" and (Table B-1) HZ limits "*a = [0.95,1.67] AU (semi-major axis)*" with footnote "*given for a solar twin. The habitable zone is scaled to √L\*/L.*" Note that the quoted limits actually correspond to the Kasting et al. (1993) "water loss" (inner edge) and maximum greenhouse (outer edge) HZ limits (see also Kopparapu et al. 2013). Following the LUVOIR & HabEx study analyses (e.g., Kopparapu et al. 2018), we do not take into account recent modeling efforts aimed at accounting for planet mass or stellar effective temperature, however such modifications could be included in the future.

   **This work**: We adopt the conservative HZ with limits (0.95, 1.67 au) for a solar twin (1 $L_{Sun}$, Teff=5772K), with semi-major axes scaled as the square root of the host star's bolometric luminosity (sqrt(L*/$L_{Sun}$); i.e. in units of the IAU nominal solar bolometric luminosity, $L_{Sun}$ = 3.828e26 W).

- **Planet-Star Brightness Ratio Limits:** The planet-star flux ratio $C = F_p/F_*$ in some passband can be calculated as:

$$C = F_p/F_* = p\, \phi(\alpha)\, (R_p/r)^2$$

(e.g., Brown 2005, Traub & Oppenheimer 2010, Burrows & Orton 2010) where *p* is the geometric albedo, $\phi(\alpha)$ is the integral phase function at phase angle $\alpha$, $R_p$ is the planet's radius, and *r* is the separation of the planet from its star. The phase angle $\alpha$ is the observer-planet-star angle (i.e., vertex at planet). Following the LUVOIR and HabEx studies, we assume the planets are on circular orbits ($r =$ semi-major axis $a$) and adopt a simple Lambertian reflectance phase function (isotropic scattering):

$$\phi(\alpha) = (\sin \alpha + (\pi - \alpha) \cos \alpha) / \pi$$

---

[3] Presumably scaling by square root of the stellar luminosity, as done elsewhere (e.g., Stark et al. 2014).

Planet-star flux ratios were calculated for two test cases: at maximum angular separation ($\alpha = 90°$) and a "gibbous" phase angle representing maximum brightness vs. separation for Lambertian spheres ($\alpha = 63.3°$, the root of the Lambertian phase angle equation between $0° < \alpha < 180°$; Brown 2005).

The starlight suppression requirement for HabEx and LUVOIR was set to a raw contrast floor of $10^{-10}$ per spatial resolution element at V band and a contrast stability requirement of $10^{-11}$ - both at the inner working angle of the system. These design goals limit the contrast at which reliable exoplanet detections and spectra can be obtained. While post-processing techniques can achieve reliable measurements well below the raw contrast level, the systematic errors determined by the contrast stability level will set a detection floor near a few times $10^{-11}$ contrast. Note that the HabEx study adopted $\Delta$mag limit of 26.5 (2.5e-11) for both its coronagraph and starshade.

**This work:** For our top tier targets, we adopted a contrast threshold level of $4 \times 10^{-11}$ ($\Delta$mag = 26.0 mags) for reliable exoplanet detections, consistent with NASA's development goals adopted for the starlight suppression technology. We appended a third tier of target stars where we considered slightly deeper limits of $2.5 \times 10^{-11}$ ($\Delta$mag = 26.5 mags).

- **Magnitude Limits:** To be a good target for spectral characterization, the planet must be bright enough so that adequate S/N can be achieved in each spectral measurement channel in a reasonable integration time. The HabEx and LUVOIR studies set a total exposure time limit of 60 days for spectral characterization, a characteristic duration of temperate planet visibility outside the inner working angle as it orbits its host star. Their preferred spectral resolution is $\lambda/\Delta\lambda = 140$, while a resolution of 70 is considered acceptable. The preferred S/N for spectra is 20 per resolution element, while 10 is considered acceptable.

  **This work:** We adopt a simplistic system throughput of 18% for the entire system including telescope, starlight suppression system, and science instrument (including detector efficiency). For $\lambda/\Delta\lambda = 70$, S/N=10, and the 60 days integration time limit, assuming an exozodi background level of three zodis and foreground local zodi background, an unobscured 6m telescope will have a limiting Rc band magnitude near 31.  Thus, we require that to be a viable target, a temperate rocky planet's apparent magnitude at the time of observation be Rc ≤ 31.

- **Inner Working Angle:**  The IWA defines the region near the star that cannot be accessed for direct imaging due to a coronagraphic mask, starshade obscuration, or an interferometric null.  It depends on the architecture of the telescope, the starlight suppression system used, and the observation wavelength.  While a nominal value could be defined by making assumptions about the design for *Habitable Worlds Observatory*, we choose instead to derive the IWA from the star list itself and Astro2020's requirement that ~100 cumulative habitable zones be surveyed by direct imaging. As shown in the main table, an IWA near ~70 mas will be needed to access ~100 cumulative habitable zones, and would need to be achieved at all wavelengths of interest for spectral characterization.   For any specific telescope and starlight suppression system architecture the IWA is typically proportional to wavelength; thus, the number of accessible HZs will strongly decrease as IWA(lambda) increases.

- **Binarity:**   The majority of stars are found in multiple systems.  Binary stars pose a complication for starlight suppression, as the light of two stars must be blocked to enable detection of the very faint exoplanet.  While excluding binary targets entirely could simplify the technical requirements for HWO, it would necessitate surveying single nearby stars across a larger volume of space - requiring either a larger telescope and/or more aggressive starlight suppression system to provide the smaller inner working angle that would be needed.  We therefore allow binary stars to be valid HWO targets, subject to constraints on their apparent separation.  Sufficiently wide binaries should be equivalent to single stars, both in terms of their lack of impact on starlight suppression performance and their likelihood of planet formation outcomes being unaffected by the companion star.

    **This work:** We adopt 10″ angular separation (almost 300 λ/D for a 6m telescope at 1.0 µm) as being sufficiently large that starlight suppression systems can perform as well for an equal-brightness binary as for a single star.  This is consistent with a primary mirror surface quality similar to that of Hubble.  We also adopt 5-10″ angular separation as the threshold for which the same quality mirror may allow sufficient performance to achieve the HWO contrast goals, pending a more detailed performance analysis of the individual targets and final determination of the telescope mirror specifications.  Finally, we adopt 3″-5″ as

the region where novel wavefront control technologies might enable coronagraphs to achieve sufficient performance for HWO to achieve its contrast goals with sufficient bandwidth. All unresolved spectroscopic binaries and binaries with separations < 3″ are excluded from the list, under the assumption that they will prove too challenging a problem for starlight suppression. Data on spectroscopic binaries was drawn from the SB9 catalog (Pourbaix et al. 2004, most recent version on Vizier at [https://cdsarc.cds.unistra.fr/viz-bin/cat/B/sb9](https://cdsarc.cds.unistra.fr/viz-bin/cat/B/sb9)) and on resolved binaries from the Washington Double Star catalog (WDS; Mason et al. 2001, most recent version on Vizier at [https://cdsarc.cds.unistra.fr/viz-bin/cat/B/wds](https://cdsarc.cds.unistra.fr/viz-bin/cat/B/wds)).

- **Search Completeness:** For nearby stars, the entire habitable zone may lie outside the inner working angle. For such targets search completeness approaching 100% can be achieved with 4-6 observations properly-spaced in time from each other, in the absence of strong aliasing between the revisit timescales and target visibility constraints. For more distant stars in the sample, only the outer part of the habitable zone may be visible outside of the inner working angle. If the obscurational completeness that can be achieved for a particular target is too small, a large number of observation epochs would be required to survey only a small fraction of the habitable zone. To avoid such targets that cannot be observed efficiently, we require that a minimum of 20% of the habitable zone be visible outside the inner working angle. For the adopted outer HZ limit (1.67 au; corresponding to the Kasting et al. (1993) and Kopparapu et al. (2013) "maximum greenhouse" case for a solar twin), this *usable outer HZ limit* corresponds to approximately 1.55 au (for a solar luminosity star). For a star to be a useful target for a nominal exo-Earth survey, this usable outer HZ limit - in angular units - would need to be larger than the IWA.

- **Exoplanet parameter cases that affect detectability:** The simple case of an Earth analog located at the Earth-equivalent instellation distance, observed at quadrature illumination, and subject to the brightness, contrast, and inner working angle limits cited above, defines only a minimal set of direct imaging targets. A temperate rocky planet can be more detectable if it has a larger radius than Earth; is observed at a gibbous illumination phase; or is located toward the outer edge of the habitable zone. To be inclusive of target stars where these other parameter cases permit detection, additional detectability scenarios need to be considered. We adopted twelve cases representing combinations of 2 planet sizes (1.0 $R_E$ Earth "twin" and 1.4 $R_E$ "super-Earth"), 2 phase angles (phase angle 90° and "gibbous" 63.3°) and 3 different instellations (1.0 au [EEID case], middle of the HZ [1.31 au], and the previously defined *usable outer HZ limit* - corresponding to the outer radius enclosing 80% of conservative HZ [1.55 au]. All of these orbital radii are then scaled by the square root of the star's bolometric luminosity relative to the Sun. The inner and outer edges of the conservative HZ themselves [0.95 au & 1.67 au] were not used because 1) the former is so similar numerically to the EEID and 2) a planet at the HZ outer edge matched to the IWA would be visible for only an infinitesimal part of its orbit. The 12 cases evaluated were:

| Case # | Phase Angle | Orbital Radius[4] | Planet Radius | Delta(mag) |
|---|---|---|---|---|
| 1 | 90° | 1.00 au | 1.0 Re | 0.00 (reference) |
| 2 | 90° | 1.00 au | 1.4 Re | -0.73 |
| 3 | 90° | 1.31 au | 1.0 Re | +0.59 |
| 4 | 90° | 1.31 au | 1.4 Re | -0.14 |
| 5 | 90° | 1.55 au | 1.0 Re | +0.95 |
| 6 | 90° | 1.55 au | 1.4 Re | +0.22 |
| 7 | 63.3° | 1.00 au | 1.0 Re | -0.64 |
| 8 | 63.3° | 1.00 au | 1.4 Re | -1.37 |
| 9 | 63.3° | 1.31 au | 1.0 Re | -0.05 |
| 10 | 63.3° | 1.31 au | 1.4 Re | -0.78 |
| 11 | 63.3° | 1.55 au | 1.0 Re | +0.31 |
| 12 | 63.3° | 1.55 au | 1.4 Re | -0.42 |

The last column lists the difference in magnitude compared to case #1 (Earth twin observed at quadrature). For now, we omit cases where the planet is smaller than 1 $R_E$, or where the planet is seen at phase angles larger than 90°, as there is no obvious circumstance where such changes would add a star to the target list that had not already been selected via the larger planet or smaller phase angle cases. The planet brightnesses for the 12 cases range over 2.32 magnitudes, with case #5 (1 $R_E$, 1.55 au,

---

[4] Where the orbital radii are scaled by the square root of the star's bolometric luminosity.

phase=90°) being the faintest, and case #8 (1.4 $R_E$, 1.00 au, phase=63.3°) being the brightest.

3. **Large Initial Input Sample Selection**

The selection of plausible target stars for the exoEarth surveys with a Decadal IR/O/UV space observatory started with nearby AFGKM stars. The original list is descended from the EPRV Working Group report (Crass, Gaudi, Leifer et al. 2021), which itself was drawn from target lists for the LUVOIR and HabEx study reports (The LUVOIR Team, 2019; Gaudi, Seager, Mennesson et al. 2020). However further effort was required to add additional stars that were bright and near enough to warrant consideration for inclusion in this provisional target star list for the *Habitable Worlds Observatory*.

Consideration of appropriate targets starts with equations predicting the reflected light exoplanet-star brightness ratio and exoplanet brightness (e.g., Brown 2005, Traub & Oppenheimer 2010):

$$C = F_p/F_* = p\,\phi(\alpha)\,(R_p/r)^2$$
$$\phi(\alpha) = (\sin\alpha + (\pi - \alpha)\cos\alpha)/\pi$$
$$\Delta mag = -2.5\log F_p/F_* = -2.5\log C$$
$$mag_p = mag_* + \Delta mag$$

where $mag_p$ is the apparent magnitude of the planet, $mag_*$ is the apparent magnitude of the star, and the other variables were previously defined earlier.

For an exoplanet receiving the Earth-equivalent insolation, the planet will have orbital radius and maximum angular separation:

$$r_{EEID} = 1\text{ au }(L_*/L_{Sun})^{½}$$
$$\theta_{EEID} = r_{EEID}/d_{pc} = r_{EEID}\,\varpi_{mas}$$

One can come up with an *approximate* distance and luminosity constraints for plausible target stars using these equations.

One can ask, *for what upper limit on luminosity can one detect an Earth twin at the EEID at maximum separation ($\alpha = 90° = \pi/2$ rad) for an adopted planet-star ratio lower limit (C = 2.5e-11)?*

$$C = 2.5\text{e-}11 = p\,\phi(\alpha = \pi/2)\,(R_p/r)^2 = p\,\pi^{-1}\,R_E^2\,r_{EEID}^{-2} = p\,\pi^{-1}\,R_E^2\,(1\text{ au})^{-2}\,(L_*/L_{Sun})^{-1}$$
$$L_*/L_{Sun} = p\,\pi^{-1}\,R_E^2\,C^{-1} = 0.2*(1/3.14159)*(6378\text{ km})^2(149597870.7\text{ km})^{-2}(2.5\text{e-}11)^{-1}$$

$$L_*/L_{Sun} = 4.6287 \Rightarrow \log_{10}(L_*/L_{Sun}) = 0.6655 \text{ dex}$$

which corresponds approximately to the bolometric luminosity of a typical F3V star with mass ~1.44 Msun. This suggests that we should probably probe all the way through the F stars and perhaps also consider at least late A-type stars. After our full analysis, the earliest spectral type among stars on the final target list was F1V (HIP 59199, Alpha Corvi A, Alchiba), and the highest luminosity star was HIP 50954 (HR 4102) which had $\log(L/L_{sun}) = 0.7128$.

One could also derive an *approximate* target distance limit as a function of stellar luminosity by equating the EEID angular separation definition ($\theta_{EEID}$; functions of stellar luminosity and distance) to an estimate of the inner working angle $\theta_{IWA}$ (functions of wavelength, aperture, and a factor related to the starlight suppression method employed), i.e.

$$r_{EEID} = 1 \text{ au } (L_*/L_{Sun})^{½}$$
$$\theta_{EEID} = r_{EEID}/d_{pc} = (L_*/L_{Sun})^{½} \, d_{pc}^{-1} = 1000 \text{ mas } (L_*/L_{Sun})^{½} \, d_{pc}^{-1}$$
$$\text{Setting } \theta_{IWA} = \theta_{EEID} = 1000 \text{ mas } (L_*/L_{Sun})^{½} \, d_{pc}^{-1}$$
$$L_*/L_{Sun} = (\theta_{IWA}/1000\text{mas})^2 \, d_{pc}^{2}$$

However, as the inner working angle will be connected to particular architecture choices (e.g., starlight suppression method, flavor of coronagraph mask, starshade, etc.), we instead decided to proceed with an iterative selection method that was technique-agnostic. Instead, our philosophy was to 1) start with an initial pool of HabEx and LUVOIR-B Hipparcos stars informed by the mission studies that was adopted for the Extreme Precision Radial Velocity Working Group Final Report[5], 2) fill in additional targets in overlapping spectral type/magnitude/distance space from various sources (e.g., SIMBAD[6]), 3) fill out volume limited samples by spectral type bin to the point at which adding more distant and fainter stars was no longer providing any additional systems where HZ exoEarths were accessible given the adopted observational constraints.

Factoring in these considerations, our input sample of stars for consideration as plausible exo-Earth survey target stars for *Habitable Worlds Observatory* was at least volume-limited by spectral class to these limits:

---

[5] https://exoplanets.nasa.gov/internal_resources/2000/ and Crass, J., Gaudi, S., Leifer, S. et al. 2021, Extreme Precision Radial Velocity Working Group Final Report, https://arxiv.org/abs/2107.14291 .
[6] https://simbad.u-strasbg.fr/simbad/ (Wenger, Ochsenbein, Egret, et al., 2000)

1) All spectral type B/A/F stars in SIMBAD out to distance 25 pc (plx > 40 mas).
2) All spectral type G stars in SIMBAD out to distance 20 pc (plx > 50 mas).
3) All spectral type K stars in SIMBAD out to distance 12 pc (plx > 83.33 mas).
4) All spectral type M dwarf stars within 5 pc (plx > 200 mas) and brighter than V=10 or G=10 in SIMBAD.

Besides stars within these limits, some additional stars just outside these limits, and companions of all of these stars, were analyzed. The total number of stars considered was approximately 800. Given the luminosity-contrast issue previously mentioned, we did not expect any A-type stars to ultimately prevail as plausible targets, however we included *all* the known A-type stars within 25 pc in our analysis just to check whether any low luminosity A stars might meet our planet detectability criteria.

For completeness, we also checked various other catalogs to see if any candidate target stars might have been missed. Besides the target stars from the LUVOIR study (for both LUVOIR-A and LUVOIR-B concepts (C. Stark, priv. Comm.; from an iteration of yield simulations) and HabEx study (HabEx Final Report 2019), we checked target lists from the recent IROUV/HWO 6-m simulations by Morgan et al. (2022), 2000s era TPF-C target lists (Brown 2005 and Stapelfeldt, priv. comm.), the recent ESA Theia concept study (Meunier et al. 2022), and the NEID GTO program (Gupta et al. 2021). To improve our chances of not missing any lower luminosity K-type stars, we also analyzed any K-type dwarfs from the XHIP catalog (Anderson & Francis 2012) and ExoCAT (Turnbull 2015) for which the calculated EEID ($a_{EEID}$ = sqrt(L/L$_{Sun}$)) in angular units ($\theta_{EEID}$ = $a_{EEID}/D_{pc}$) was greater than 42 mas.

Stars found to be binaries in the literature were split into multiple entries when there was sufficient information to derive parameters for individual components. Additional faint companions to these stars were tracked to to help with characterizing multiple stars systems, even if these companions would have normally failed our selection criteria.

## 4. Final Criteria Applied to Select the Star List

The goal of this analysis was to identify of order ~100 stellar targets which are likely to be among those that will enable the future *Habitable Worlds Observatory* to fulfill its Astro 2020 Decadal Survey goal of studying ~25 potentially habitable worlds. Using the parent sample and evaluating the brightnesses and angular separations for the 12 nominal exoplanet cases previously described for each star, we counted the number of cases where the exoplanet satisfied three constraints (angular separation > IWA, planet-star brightness ratio above a minimum threshold, and exoplanet apparent Rc magnitude brighter than some threshold). The targets were iteratively ranked by the number of exoplanet cases that exceeded the thresholds (down to >=6 cases, i.e. >=50% of cases) and by Earth twin apparent magnitude. Targets were assigned to one of three tiers (A, B, or C; see table below), or rejected outright.

The selection criteria were applied to the input sample, initially selecting stars that satisfied all of the criteria and working our way from 12 cases satisfied down to 6. Along the way, the cases that had issues with binarity or disks were relegated to tiers B or C. 7 targets had all 12 hypothetical exoplanets satisfy the Tier A criteria - i.e. they were consistently the "best" targets (HIP 104214 (61 Cyg A), HIP 108870 (eps Ind), HIP 19849 A (40 Eri A), HIP 96100 (sig Dra), HIP 61317 (bet CVn), HIP 15457 (kap1 Cet), HIP 57443 A (HR 4523)). Setting the exoplanet brightness limit to Rc = 30.5 and exoplanet-star brightness minimum limit to 4e-11, we were able to select 47 targets with no known disks, and which were either single or binaries with companions at >10″ separations, by decreasing the IWA down to 83 mas[7]. As this provided roughly half of the anticipated number of needed targets, we considered these to be the best "half" of the targets needed for an HWO exo-Earth survey and considered them "**Tier A**".

To add additional stars, the IWA was decreased another 14% down to 72 mas (in order to provide selection of another ~50 targets), and a slightly fainter planet magnitude limit was considered ($R_c$ = 31.0), while maintaining the same planet-star brightness limit (4e-11). Slightly looser criteria were allowed for disks and binaries - i.e. stars with cold Kuiper Belt disks were allowed if they were very optically thin ($L_{IR}/L_* < 10^{-4}$), and binaries with separations as small as 5″ were allowed. Hence the stars in this 2nd tier of targets ("**Tier B**") all have one or more issues which make them less desirable than those in Tier A, i.e., their HZ planets would be fainter, or closer-in, or having lower planet-star brightness ratioes, or they may have an optically thin cold disk, or they may be a binary requiring further analysis to see whether they could still be good targets. Two stars with known giant planets in, or near, their habitable zones (HIP 3093 = 54 Psc A and HIP 86796 = Mu Arae = Cervantes) were retained as Tier B stars, but none of the Tier A targets are known to have such potentially disruptive giant planets. We note that there is some observational evidence for a correlation between detectable cold dust disks and detectable warm exozodiacal dust disks among nearby stars (Ertel et al. 2020), so the cold dust systems may be more likely to have elevated exozodi background levels in the UV/visible/near-IR for HWO observations. We anticipate that a significant number of these Tier B targets may ultimately prove unsuitable for exoEarth surveys after further analysis. The Tier A and B samples together get us to "near" 100 systems (47 + 51 = 98 stars), and approximately ~93 cumulative habitable zones - counting just the fractions of the HZ accessible outside of the IWA constraints in the table below (72 mas after building Tiers A and B). We add a third tier (**Tier C**) of slightly less optimal targets - with the expectation that further characterization of these systems may lead to some of them eventually being elevated to help fulfill the 100 cumulative HZ sample.

---

[7] For comparison, 83 mas corresponds is the IWA predicted for wavelength 1.0 μm (continuum level just redward of an important habitability signature – the water absorption feature at 940nm) for an aperture of 6 meters for a vector vortex charge-6 coronagraph (adopted by HabEx and LUVOIR) that has an inner working angle of 2.4 λ/D.

The third tier of targets (**Tier C**) allowed for an even slightly smaller IWA than Tier B (another 10% smaller, down to 65 mas), keeping the same exoplanet brightness limit ($R_c$ = 31.0), but allowing for a slightly more aggressive exoplanet-star brightness ratio limit (2.5e-11). Tier C also included binaries and disks which at first glance might prove even more problematic for exo-Earth surveys, but worth further investigation and analysis. For Tier C we also allowed binaries down to 3″ separation, and those with any known dust disks (i.e., even those of higher optical depth than allowed in Tier B). Tier C contained an additional 66 stars. This brought the total number of Tier A, B, and C target stars to 164.

| Parameter | Tier A | Tier B | Tier C |
|---|---|---|---|
| **IWA constraint** | 83 mas | 72 mas | 65 mas |
| **Exoplanet brightness limit (Rc)** | 30.5 mag | 31.0 mag | 31.0 mag |
| **Exoplanet-star Brightness ratio limit** | 4e-11 | 4e-11 | 2.5e-11 |
| **Disk criterion** | No known dust disks of any kind | No disk, or KB disks OK if $L_{disk}/L_* <= 10^{-4}$ | All disks OK, even if $L_{disk}/L_* >= 10^{-4}$ or detected HZ warm dust disk |
| **Treatment of binaries** | Single or binary companion > 10″ sep | Single or binary companion 5″ - 10″ sep | Single or binary companion 3″ - 5″ sep |
| **Number of Stars** | 47 | 51 | 66 |

All target stars that had a binary companion at separation less than 3″, including unresolved spectroscopic binaries, were discarded for this initial selection effort.

Note that after the initial analysis, further examination of the literature identified three stars that were otherwise good targets but were initially flagged as close binaries and omitted. However, closer examination of the literature suggests that the early reports of binarity for these systems were most likely to be spurious. These stars were **HIP 4151, HIP 23835, HIP 86736**. All 3 were added to Tier "C" as the 64th, 65th, and 66th entries. Despite their indications of binarity in the literature, the preponderance of measurements over the past few decades suggests that these stars are more likely to be single. Further observations would be useful to test whether these systems are indeed single.

## 5. Characteristics of Selected Stars

Figure 1 shows the distributions of distance versus luminosity for the selected target stars by tier. Figure 2 shows the HR diagram for the selected stars, and Figure 3 shows histograms for some of the stellar properties of the target stars (apparent V magnitude, effective temperature, mass, and metallicity [Fe/H]). The target stars are listed in the Table in Appendix A, sorted by RA, and flagged by tier. The sample tiers break down by spectral type as follows:

| Sample | F | G | K | M |
|---|---|---|---|---|
| Tier A | 14 | 15 | 17 | 1 |
| Tier B | 15 | 23 | 11 | 2 |
| Tier C | 37 | 17 | 12 | 0 |
| Total (A+B+C) | 66 | 55 | 40 | 3 |

No A-type stars were selected given our selection criteria, despite the input sample being complete for the 33 A-type stars within d=25 pc. The lone B-type star within 25 pc (Regulus, HIP 49669) also did not make the cut. The high luminosities of the B/A-type stars push their habitable zones to larger orbital radii (making them more accessible in terms of IWA constraints), however the larger orbital radii for temperate exoplanets lead to less of the star's light being reflected by the planets for a given size and albedo, resulting in low planet-star brightness ratios and faint planet magnitudes (below our selection thresholds). All of the A-type main sequence stars within 25 pc have isochronal ages of <1.0 Gyr, consistent with the nuclear lifetimes of stars with masses ~1.6-3.0 Msun (David & Hillenbrand 2015).

The three M dwarfs selected are:
HIP 105090 = Lacaille 8760 (M0V, V=6.65, Rc=5.77, EEID=73mas, d=4.0pc; tier A)
HIP 114046 = Lacaille 9352 (M1V, V=7.33, Rc=6.36, EEID=58mas, d=3.3pc; tier B)
HIP 54035 = Lalande 21185 (M2V, V=7.42, Rc=6.40, EEID=55mas, d=2.6pc; tier B)

These three stars are recognized as the three brightest M dwarfs in the sky[8]. All other known M dwarfs have EEID < 44 mas and are ~>0.5 mag fainter in the V and Rc bands than the faintest of these three (Lalande 21185). The next best M dwarfs, in order of EEID would have been HIP 25878 (GJ 205; EEID = 44 mas), HIP 45343 (GJ 338A; EEID=42 mas), HIP 73182 (GJ 570B; EEID=41 mas), and HIP 29295 (GJ 229; EEID=40 mas). However, none of these stars had more than 2 of the fiducial exoplanet visibility cases detectable (among the 12 cases tested), and did not satisfy our selection criteria. An additional M dwarf would not have been added unless the IWA threshold was lowered to 55 mas (which would have been HIP 25878 = GJ 205).

The target list is dominated by main sequence dwarf stars, and contains only a few subgiants, and no giants. Surface gravities for the stars from the literature range from log(g) = 3.78 to 4.88. Only 4 target stars have low surface gravities of log(g) < 4, with the lowest gravity star being the K0+IV subgiant Rana (del Eri, HIP 17378) with log(g)=3.78. In Figure 3, we plot histograms for some of the stellar parameters for the target stars (apparent V magnitude, effective temperature, stellar mass, and metallicity) based on fiducial literature values, with the names of some of the stars with extreme values IDed.

---

[8] See 2002 article by Ken Croswell http://kencroswell.com/thebrightestreddwarf.html .

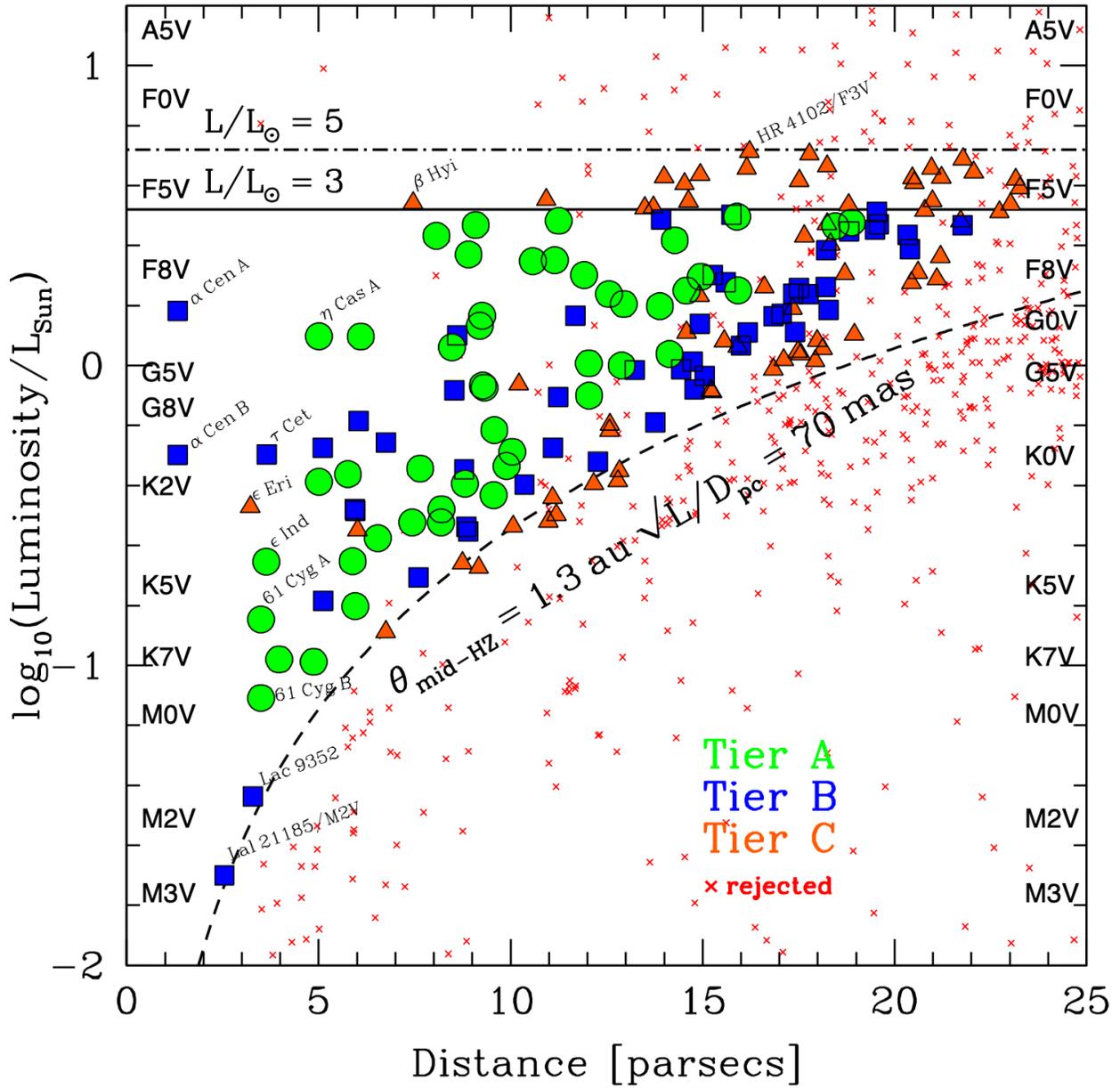

**Figure 1:** Distance versus bolometric luminosity for the target stars (Tier A stars are large green circles, Tier B stars are blue squares, Tier C stars are small orange triangles). Stars that were characterized and tested but not selected are red Xs. Characteristic lines are plotted showing stellar luminosities of 3 Lsun (solid) and 5 Lsun (dash-dot), which roughly define upper limits for the Tiers A+B and C samples, respectively. The long-dash curve *roughly* traces the lower envelope of target stars, with the line drawn equating to where the middle of the habitable zone (1.31au X sqrt(luminosity) would appear at angular separation 70mas. The line itself was *not* used to select stars, but visibly approximately traces the lower selection boundary.

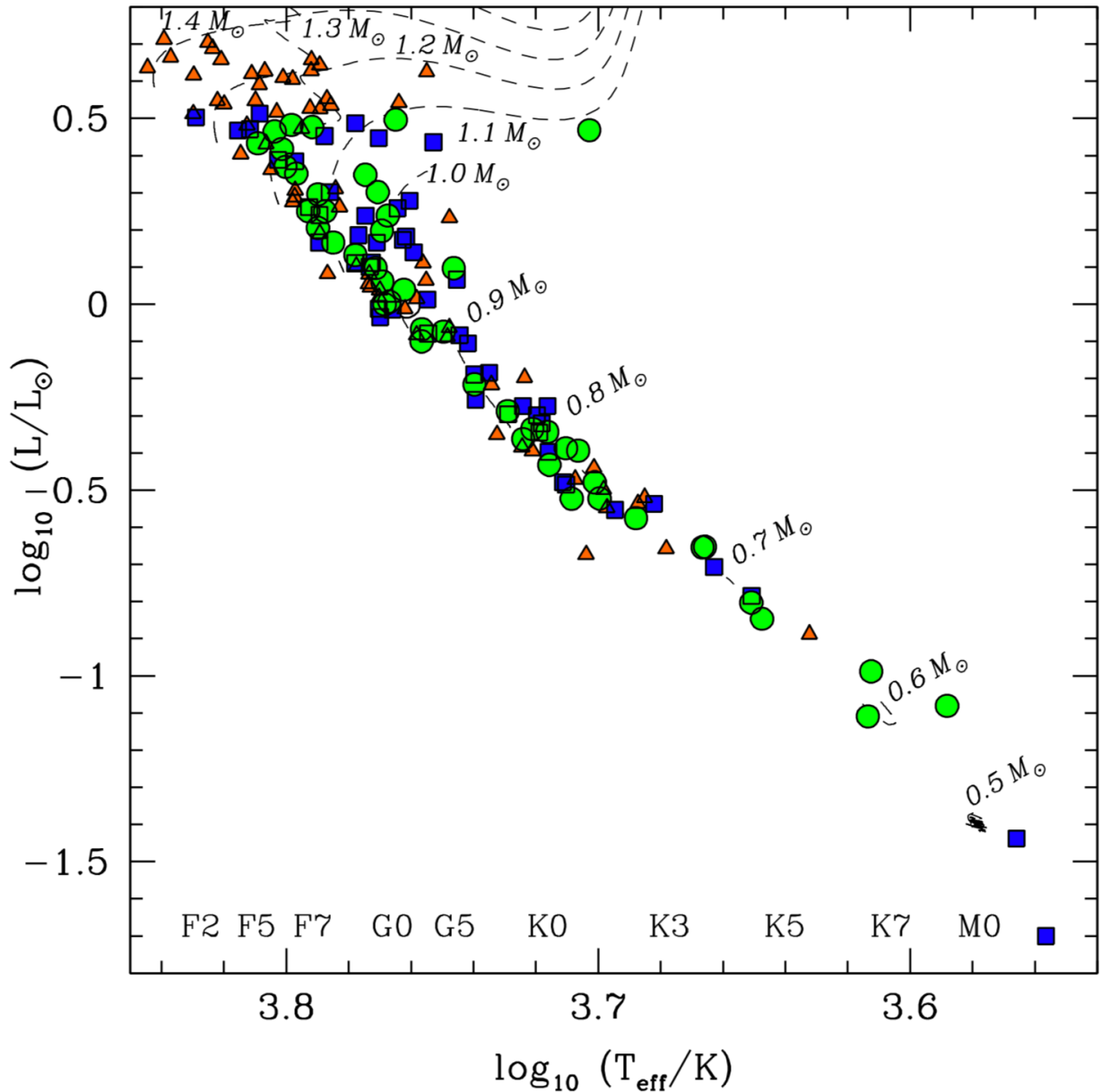

**Figure 2**: HR diagram - effective temperature versus bolometric luminosity - for the selected target stars (Tier A stars are large green circles, Tier B stars are blue squares, Tier C stars are small orange triangles). MIST evolutionary tracks for 0.5 to 1.4 solar masses (in steps of 0.1 Msun) from Choi et al. (2016) for solar composition between ages 100 Myr and 10 Gyr are shown as dashed lines.

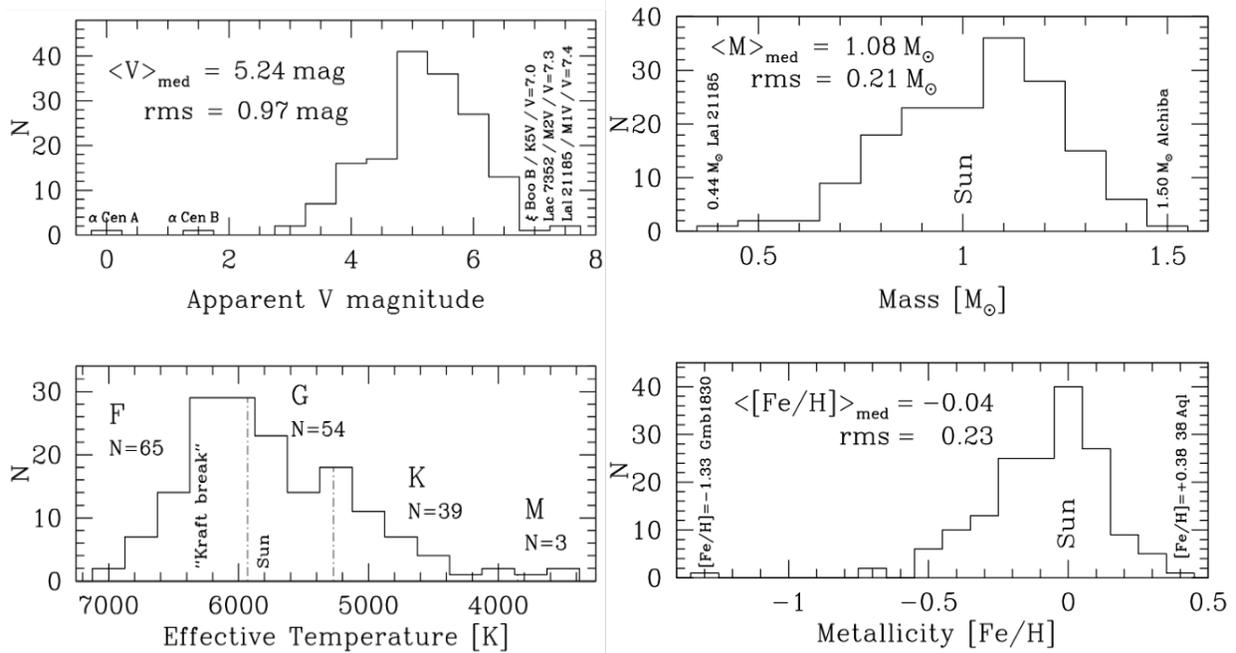

**Figure 3**: Histograms of the distributions of stellar parameters for the targets (including all three tiers A, B, and C): apparent V magnitudes (*upper left*), effective temperatures (*lower left*), fiducial masses (*upper right*), and metallicities (*lower right*).

In addition to the list of target stars in Appendix A, we include some important lists of stars in additional appendices. In Appendix B, we list stars from the HabEx and LUVOIR-B target lists that were not included in the ExEP mission target list. In Appendix C, we list the ExEP target stars that were not included in the HabEx or LUVOIR-B target lists. Appendix D lists target stars in Tiers B and C with separations ~3″-10″ whose orbital characteristics and system parameters should be investigated further to assess whether they should be retained as viable targets. Appendix E contains stars (mix of Tiers A, B, and C) that have been previously reported to be binaries in the literature at one point, but for which subsequent observations have not confirmed the binarity or judged it to be spurious. Further observations and analysis is probably warranted to make sure that they truly are spurious or whether some elusive companion is actually present. Appendix F contains a description of the columns in the target list table.

## 6. Concluding Comments

The target list we have constructed is optimal in the sense that it provides close to 100 cumulative habitable zones that are most-widely-separated from their host stars. This choice maximizes the inner working angle required for *Habitable Worlds Observatory* to achieve this goal, allowing for the smallest telescope and/or least aggressive starlight suppression system. However, our analysis has not included consideration of how much mission time would be needed to conduct a blind survey of all the targets and to do the spectral characterization of the temperate rocky planets that are found. It is possible that a brighter overall target set could be obtained by requiring a more aggressive inner working angle, contrast, and/or system stability, thus achieving Astro2020's survey goal in a shorter mission time. The HWO maturation process will need to consider this question in detail, balancing mission lifetime requirements against starlight suppression requirements.

Our target analysis shows that the number of stars accessible out to 1 $\mu$m with the vector vortex charge-6 coronagraph design and a 6m aperture will fall short of the Decadal 100 HZ requirement, by a factor of two. Either a more capable coronagraph, a telescope aperture close to 8m, or a starshade will be required to survey a 100 cumulative HZ sample out to 1 $\mu$m. Alternatively, the long wavelength spectral cutoff could be scaled back, at the cost of losing coverage of the 0.94 $\mu$m water band in a significant fraction of the targets.

## 7. Acknowledgements

We would like to thank Bertrand Mennesson, Rhonda Morgan, Josh Pepper, Aki Roberge, Dmitry Savransky, Chris Stark, and Maggie Turnbull for their review and comments on the data table and documentation. We thank Jessie Christiansen, David Ciardi, and Anjali Tripathi for discussions and feedback on the documentation and formatting. We'd like to thank members of the EPRV Working Group, ExoPAG SAG 22, and HabEx and LUVOIR concept study teams, who provided input, analysis, and valuable reports that factored into refining this provisional target list. This research was carried out at the Jet Propulsion Laboratory, California Institute of Technology, under a contract with the National Aeronautics and Space Administration (80NM0018D0004). Feedback may be directed to Eric Mamajek (mamajek@jpl.nasa.gov) and Karl Stapelfeldt (karl.r.stapelfeldt@jpl.nasa.gov).## 8. References

- Abt, H.A. & Levy, S.G., 1976, ApJS, 30, 273
- Anderson, E. & Francis, Ch., 2012, Astronomy Letters, 38, 331
- Bakos, G.A., 1964, SAO Special Report #162 (1964)
- Bessell, M.S., 1990, A&AS, 83, 357
- Bessell, M.S., Castelli, F., & Plez, B. 1998, A&A, 333, 231
- Brown, 2005, ApJ, 624, 1010
- Burrows, A. & Orton, G., 2010, Exoplanets, edited by S. Seager. Tucson, AZ: University of Arizona Press, 2010, 526 pp. ISBN 978-0-8165-2945-2., p.419-44021

- Riello, M., De Angeli, F., Evans, D.W., et al., 2021, A&A, 649, A3
- Skiff, B.A., 2014, VizieR Online Data Catalog: Catalogue of Stellar Spectral Classifications, B/mk
- Stark, C., Roberge, A., Mandell, A., Robinson, T.D., 2014, ApJ, 795, 122
- Stassun, K.G., Oelkers, R.J., Paegert, M., et al., 2019, AJ, 158, 138
- Wenger, M., Ochsenbein, F., Egret, D., et al., 2000, A&AS, 143, 9
- Traub, W.A., & Oppenheimer, B.R., 2010, Exoplanets, edited by S. Seager. Tucson, AZ: University of Arizona Press, 2010, 526 pp. ISBN 978-0-8165-2945-2., p.111-156
- Turnbull, M., 2015, arXiv:1510.01731

**Appendix A: Provisional NASA ExEP Target Star List for Precursor and Preparatory Science for *Habitable Worlds Observatory* (2023)**

Table A contains the provisional ExEP list of high priority target stars as of January 2023 for precursor and preparatory science to inform design of *Habitable Worlds Observatory* and encourage research to facilitate future surveys for exo-Earths to fulfill a key recommendation of the Astro2020 Decadal Survey. Table A columns include stellar designations (HIP, HD, IAU proper name or common name/ID), apparent V magnitude, distance in parsecs, spectral type, and tier category (A, B, or C). For clarification in cases of multiple systems, component identifiers (e.g., A, B) are listed with the IDs - *even if the target star itself is often known in the literature by the ID without the component identifier*. For example, the A component in the 54 Piscium binary is listed below with IDs "54 Piscium A", "HD 3651 A", and "HIP 3093 A", but appears in SIMBAD with identifiers "54 Psc", "HD 3651", and "HIP 3093". Component IDs follow usage in the Washington Double Star Catalog (WDS). The spectral types are abbreviated in some cases when additional specifiers are provided from literature sources (often related to metallicity or activity), and references are provided in the separate table file. Spectral types were found from queries of SIMBAD and the Skiff (2014) compendium of spectral classifications, and the vast majority of quoted types come from four sources: Keenan & McNeil (1989), Gray et al. (2001), and the NStars/Space Interferometry Mission Preparatory Science Program spectral classification surveys of Gray et al. (2003, 2006).

**Table A: Provisional NASA ExEP target star list for precursor and preparatory science for *Habitable Worlds Observatory* (2023)**

| ID(HIP) | ID(HD) | Common Name | V(mag) | Dist(pc) | SpT | Tier |
|---|---|---|---|---|---|---|
| HIP 544 | HD 166 | V439 Andromedae | 6.09 | 13.77 | G8V | B |
| HIP 910 | HD 693 | 6 Ceti | 4.90 | 18.89 | F8V | A |
| HIP 950 | HD 739 | Theta Sculptoris | 5.24 | 21.72 | F5V | C |
| HIP 1599 | HD 1581 | Zeta Tucanae | 4.22 | 8.61 | F9.5V | B |
| HIP 2021 | HD 2151 | Beta Hydri | 2.82 | 7.46 | G0V | C |
| HIP 3093 A | HD 3651 A | 54 Piscium A | 5.86 | 11.11 | K0.5V | B |
| HIP 3583 A | HD 4391 | HD 4391 | 5.80 | 15.05 | G5V | B |
| HIP 3765 | HD 4628 | HD 4628 | 5.73 | 7.44 | K2V | A |
| HIP 3821 A | HD 4614 A | Achird | 3.44 | 5.01 | F9V | A |
| HIP 3909 | HD 4813 | 19 Ceti | 5.18 | 15.92 | F7V | A |
| HIP 4151 | HD 5015 | HD 5015 | 4.80 | 18.80 | F8V | C |
| HIP 5862 | HD 7570 | Nu Phoenicis | 4.97 | 15.26 | F9V | B |
| HIP 5896 A | HD 7788 A | Kappa Tucanae A | 4.91 | 23.26 | F5V | C |
| HIP 7513 A | HD 9826 A | Titawin | 4.10 | 13.49 | F8V | C |
| HIP 7751 B | HD 10361 | p Eridani B | 5.88 | 8.20 | K2V | A |
| HIP 7751 A | HD 10360 | p Eridani A | 5.76 | 8.19 | K2V | A |
| HIP 7978 | HD 10647 | q1 Eridani | 5.52 | 17.35 | F9V | C |
| HIP 7981 | HD 10476 | 107 Piscium | 5.24 | 7.64 | K1V | A |
| HIP 8102 | HD 10700 | Tau Ceti | 3.50 | 3.65 | G8V | B |
| HIP 8362 | HD 10780 | V987 Cassiopeiae | 5.63 | 10.04 | G9V | A |
| HIP 10798 | HD 14412 | HD 14412 | 6.34 | 12.83 | G8V | C |
| HIP 12653 | HD 17051 | Iota Horologii | 5.40 | 17.36 | F9V | B |
| HIP 12777 A | HD 16895 A | Theta Persei A | 4.10 | 11.15 | F7V | A |
| HIP 12843 | HD 17206 | 1 Eridani | 4.47 | 14.28 | F7V | A |
| HIP 13402 | HD 17925 | EP Eridani | 6.04 | 10.36 | K1.5V | B |
| HIP 14632 | HD 19373 | Iota Persei | 4.05 | 10.58 | G0V | A |
| HIP 14879 A | HD 20010 A | Dalim | 3.80 | 14.00 | F6V | C |
| HIP 15330 | HD 20766 | Zeta1 Reticuli | 5.51 | 12.04 | G2IV | A |
| HIP 15371 | HD 20807 | Zeta2 Reticuli | 5.23 | 12.04 | G1V | A |
| HIP 15457 | HD 20630 | Kappa1 Ceti | 4.85 | 9.28 | G5V | A |
| HIP 15510 | HD 20794 | 82 Eridani | 4.26 | 6.04 | G6V | B |
| HIP 16245 A | HD 22001 A | Kappa Reticuli A | 4.70 | 21.78 | F3V | C |
| HIP 16537 | HD 22049 | Ran | 3.72 | 3.22 | K2V | C |
| HIP 16852 | HD 22484 | 10 Tauri | 4.29 | 13.92 | F9IV-V | B |
| HIP 17378 | HD 23249 | Rana | 3.54 | 9.09 | K0+IV | A |
| HIP 17651 | HD 23754 | 27 Eridani | 4.21 | 17.78 | F5IV-V | C |

| HIP 18859 | HD 25457 | HD 25457 | 5.36 | 18.71 | F7V | C |
| --- | --- | --- | --- | --- | --- | --- |
| HIP 19335 | HD 25998 | 50 Persei | 5.52 | 21.19 | F8V | C |
| HIP 19849 A | HD 26965 A | Keid | 4.42 | 5.01 | K0.5V | A |
| HIP 22263 | HD 30495 | 58 Eridani | 5.49 | 13.24 | G1.5V | B |
| HIP 22449 | HD 30652 | Tabit | 3.18 | 8.07 | F6V | A |
| HIP 23311 | HD 32147 | HD 32147 | 6.20 | 8.84 | K3+V | B |
| HIP 23693 | HD 33262 A | Zeta Doradus A | 4.70 | 11.69 | F9V | B |
| HIP 23835 | HD 32923 | 104 Tauri | 4.92 | 15.92 | G1V | C |
| HIP 24813 | HD 34411 | Lambda Aurigae | 4.71 | 12.56 | G1.5V | A |
| HIP 25110 | HD 33564 | HD 33564 | 5.08 | 20.79 | F7V | C |
| HIP 25278 | HD 35296 | 111 Tauri | 5.01 | 14.58 | F8V | A |
| HIP 26394 | HD 39091 | Pi Mensae | 5.67 | 18.29 | G0V | B |
| HIP 26779 | HD 37394 | V538 Aurigae | 6.20 | 12.27 | K1V | B |
| HIP 27072 B | HD 38392 | Gamma Leporis B | 6.14 | 8.89 | K2.5V | B |
| HIP 27072 A | HD 38393 | Gamma Leporis A | 3.60 | 8.91 | F6.5V | A |
| HIP 27435 | HD 38858 | HD 38858 | 5.97 | 15.21 | G2V | C |
| HIP 29271 A | HD 43834 A | Alpha Mensae A | 5.08 | 10.21 | G7V | C |
| HIP 29650 | HD 43042 | 71 Orionis | 5.20 | 21.78 | F5.5IV-V | B |
| HIP 29800 | HD 43386 | 74 Orionis | 5.04 | 19.59 | F5V | B |
| HIP 32439 A | HD 46588 A | HD 46588 A | 5.44 | 18.20 | F8V | B |
| HIP 32480 | HD 48682 | 56 Aurigae | 5.25 | 16.61 | F9V | C |
| HIP 32984 A | HD 50281 | HD 50281 | 6.56 | 8.74 | K3.5V | C |
| HIP 33277 | HD 50692 | 37 Geminorum | 5.76 | 17.40 | G0V | B |
| HIP 34065 | HD 53705 | HD 53705 | 5.56 | 17.06 | G1.5V | B |
| HIP 35136 | HD 55575 | HD 55575 | 5.56 | 16.85 | F9V | B |
| HIP 36439 | HD 58855 | 22 Lyncis | 5.35 | 20.40 | F6V | B |
| HIP 38423 A | HD 64379 | 212 Puppis A | 5.09 | 18.33 | F5V | C |
| HIP 38908 A | HD 65907 A | HD 65907 A | 5.59 | 16.17 | F9.5V | B |
| HIP 40693 | HD 69830 | HD 69830 | 5.95 | 12.58 | G8+V | C |
| HIP 40843 | HD 69897 | Chi Cancri | 5.13 | 18.22 | F6V | B |
| HIP 41926 | HD 72673 | HD 72673 | 6.38 | 12.16 | K1V | C |
| HIP 42438 | HD 72905 | 3 Ursae Majoris A | 5.63 | 14.44 | G0.5V | B |
| HIP 42808 | HD 74576 | HD 74576 | 6.56 | 11.19 | K2.5V | C |
| HIP 43587 A | HD 75732 A | Copernicus | 5.96 | 12.59 | K0IV-V | C |
| HIP 43726 | HD 76151 | HD 76151 | 6.01 | 16.85 | G2V | C |
| HIP 44897 | HD 78366 | HD 78366 | 5.96 | 18.95 | G0IV-V | C |
| HIP 45038 A | HD 78154 A | 13 Ursae Majoris A | 4.81 | 20.52 | F7V | C |
| HIP 47080 A | HD 82885 A | 11 Leo Minoris A | 5.40 | 11.23 | G9-IV-V | B |
| HIP 47592 | HD 84117 | HD 84117 | 4.91 | 14.95 | F9V | A |
| HIP 48113 | HD 84737 | HD 84737 | 5.09 | 18.82 | G0V | B |
| HIP 49081 A | HD 86728 A | 20 Leo Minoris A | 5.38 | 14.93 | G4IV | B |

| HIP 49908 | HD 88230 | Groombridge 1618 | 6.55 | 4.87 | K7V | A |
| HIP 50564 | HD 89449 | 40 Leonis | 4.79 | 21.22 | F6IV-V | C |
| HIP 50954 | HD 90589 | HD 90589 | 3.99 | 16.22 | F3V | C |
| HIP 51459 A | HD 90839 | 36 Ursae Majoris A | 4.82 | 12.95 | F8V | A |
| HIP 51502 A | HD 90089 A | HD 90089 A | 5.25 | 22.73 | F4V | C |
| HIP 51523 | HD 91324 | HD 91324 | 4.90 | 22.06 | F9V | C |
| HIP 53721 | HD 95128 | Chalawan | 5.04 | 13.89 | G1.5IV-V | A |
| HIP 54035 | HD 95735 | Lalande 21185 | 7.42 | 2.55 | M2V | B |
| HIP 56452 A | HD 100623 A | 20 Crateris A | 5.96 | 9.56 | K0-V | A |
| HIP 56997 | HD 101501 | 61 Ursae Majoris | 5.31 | 9.58 | G8V | A |
| HIP 57443 A | HD 102365 | HD 102365 | 4.89 | 9.32 | G2V | A |
| HIP 57757 | HD 102870 | Zavijava | 3.60 | 10.93 | F9V | C |
| HIP 57939 | HD 103095 | Groombridge 1830 | 6.43 | 9.17 | K1V | C |
| HIP 59199 A | HD 105452 A | Alchiba | 4.03 | 14.94 | F1V | C |
| HIP 61174 | HD 109085 | Eta Corvi | 4.30 | 18.24 | F2V | C |
| HIP 61317 | HD 109358 | Chara | 4.26 | 8.47 | G0V | A |
| HIP 62207 | HD 110897 | 10 Canum Venaticorum | 5.96 | 17.56 | F9V | C |
| HIP 64394 | HD 114710 | Beta Comae Berenices | 4.23 | 9.20 | F9.5V | A |
| HIP 64408 | HD 114613 | HD 114613 | 4.85 | 20.46 | G4IV | C |
| HIP 64583 A | HD 114837 A | HD 114837 A | 4.91 | 18.24 | F6V | C |
| HIP 64797 A | HD 115404 A | HD 115404 A | 6.55 | 10.99 | K2.5V | C |
| HIP 64924 | HD 115617 | 61 Virginis | 4.74 | 8.53 | G6.5V | B |
| HIP 68184 | HD 122064 | HR 5256 | 6.49 | 10.07 | K3V | C |
| HIP 69965 A | HD 125276 A | HD 125276 A | 5.87 | 17.99 | F9V | C |
| HIP 70497 A | HD 126660 A | Theta Bootis A | 4.05 | 14.53 | F7V | C |
| HIP 71284 | HD 128167 | Sigma Bootis | 4.47 | 15.76 | F4V | B |
| HIP 71681 | HD 128621 | Toliman | 1.35 | 1.33 | K1V | B |
| HIP 71683 | HD 128620 | Rigil Kentaurus | 0.00 | 1.33 | G2V | B |
| HIP 72659 B | HD 131156 B | Xi Bootis B | 6.98 | 6.75 | K5V | C |
| HIP 72659 A | HD 131156 A | Xi Bootis A | 4.54 | 6.75 | G8-V | B |
| HIP 73184 | HD 131977 | GJ 570A | 5.72 | 5.89 | K4V | A |
| HIP 73996 | HD 134083 | 45 Bootis | 4.94 | 19.54 | F5V | B |
| HIP 75181 | HD 136352 | Nu2 Lupi | 5.66 | 14.74 | G2.5V | B |
| HIP 77052 A | HD 140538 A | Psi Serpentis A | 5.87 | 14.79 | G5V | B |
| HIP 77257 | HD 141004 | Lambda Serpentis | 4.42 | 11.92 | G0-V | A |
| HIP 77358 A | HD 140901 A | HD 140901 A | 6.01 | 15.25 | G7IV-V | C |
| HIP 77760 | HD 142373 | Chi Herculis | 4.61 | 15.90 | G0V | A |
| HIP 78072 | HD 142860 | Gamma Serpentis | 3.84 | 11.25 | F6V | A |
| HIP 78459 | HD 143761 | Rho Coronae Borealis | 5.41 | 17.51 | G0IV | B |
| HIP 79672 | HD 146233 | 18 Scorpii | 5.50 | 14.14 | G3+IV | A |
| HIP 80337 | HD 147513 | HD 147513 | 5.37 | 12.89 | G1V | A |

| HIP 81300 | HD 149661 | 12 Ophiuchi | 5.76 | 9.89 | K0V | A |
| HIP 84405 A | HD 155886 | Guniibuu | 5.07 | 5.95 | K1V | B |
| HIP 84405 B | HD 155885 | Guniibuu B | 5.11 | 5.95 | K1V | B |
| HIP 84478 | HD 156026 | Guniibuu C | 6.30 | 5.95 | K5V | A |
| HIP 84720 A | HD 156274 A | 41 Arae A | 5.47 | 8.79 | G9V | B |
| HIP 84862 | HD 157214 | 72 Herculis | 5.39 | 14.59 | G0V | C |
| HIP 84893 A | HD 156897 A | Aggia | 4.39 | 17.52 | F2V | C |
| HIP 85235 | HD 158633 | HD 158633 | 6.44 | 12.79 | K0V | C |
| HIP 86486 | HD 160032 | Lambda Arae | 4.76 | 20.96 | F4V | C |
| HIP 86736 | HD 160915 | 58 Ophiuchi | 4.86 | 17.65 | F5V | C |
| HIP 86796 | HD 160691 | Cervantes | 5.12 | 15.60 | G3V | B |
| HIP 88601 A | HD 165341 A | 70 Ophiuchi A | 4.22 | 5.11 | K0-V | B |
| HIP 88601 B | HD 165341 B | 70 Ophiuchi B | 6.06 | 5.12 | K4V | B |
| HIP 88694 A | HD 165185 | HD 165185 | 5.95 | 17.11 | G0V | C |
| HIP 88972 | HD 166620 | HD 166620 | 6.38 | 11.10 | K2V | C |
| HIP 89042 | HD 165499 | Iota Pavonis | 5.47 | 17.75 | G0V | B |
| HIP 89348 | HD 168151 | 36 Draconis | 4.99 | 23.16 | F5V | C |
| HIP 95447 | HD 182572 | 31 Aquilae | 5.17 | 14.92 | G7IV-V | C |
| HIP 96100 | HD 185144 | Alsafi | 4.67 | 5.76 | K0V | A |
| HIP 97295 A | HD 187013 | 17 Cygni A | 5.01 | 20.99 | F5.5IV-V | C |
| HIP 97675 A | HD 187691 A | Omicron Aquilae A | 5.12 | 19.49 | F8V | B |
| HIP 98767 | HD 190360 | HD 190360 | 5.75 | 16.00 | G7V | B |
| HIP 98959 | HD 189567 | HD 189567 | 6.07 | 17.93 | G2V | C |
| HIP 99240 | HD 190248 | Delta Pavonis | 3.56 | 6.10 | G8IV-V | A |
| HIP 99461 A | HD 191408 A | HD 191408 A | 5.30 | 6.01 | K2.5V | C |
| HIP 99825 | HD 192310 | HD 192310 | 5.73 | 8.81 | K2+V | A |
| HIP 100017 | HD 193664 | HD 193664 | 5.92 | 17.48 | G0V | C |
| HIP 102485 | HD 197692 | Psi Capricorni | 4.14 | 14.63 | F5V | C |
| HIP 103389 | HD 199260 | HD 199260 | 5.71 | 21.10 | F6V | C |
| HIP 104214 | HD 201091 | 61 Cygni A | 5.21 | 3.50 | K5V | A |
| HIP 104217 | HD 201092 | 61 Cygni B | 6.04 | 3.50 | K7V | A |
| HIP 105090 | HD 202560 | Lacaille 8760 | 6.65 | 3.97 | M0V | A |
| HIP 105858 | HD 203608 | Gamma Pavonis | 4.23 | 9.26 | F9V | A |
| HIP 107350 | HD 206860 | HN Pegasi | 5.94 | 18.13 | G0IV-V | C |
| HIP 107649 | HD 207129 | HD 207129 | 5.58 | 15.56 | G0V | C |
| HIP 108870 | HD 209100 | Epsilon Indi | 4.67 | 3.64 | K4V | A |
| HIP 109422 | HD 210302 | Tau Piscis Austrini | 4.94 | 18.46 | F6V | A |
| HIP 110649 A | HD 212330 A | HD 212330 A | 5.32 | 20.34 | G2IV-V | B |
| HIP 111449 A | HD 213845 A | Upsilon Aquarii A | 5.21 | 23.02 | F5V | C |
| HIP 112447 A | HD 215648 A | Xi Pegasi A | 4.20 | 16.15 | F6V | C |
| HIP 113283 | HD 216803 | Fomalhaut B | 6.45 | 7.60 | K4V | B |

| HIP 114046 | HD 217987 | Lacaille 9352 | 7.33 | 3.29 | M1V | B |
| HIP 114622 | HD 219134 | HD 219134 | 5.54 | 6.54 | K3V | A |
| HIP 114924 | HD 219623 | HD 219623 | 5.58 | 20.61 | F8V | C |
| HIP 114948 | HD 219482 | HD 219482 | 5.66 | 20.44 | F6V | C |
| HIP 116771 | HD 222368 | Iota Piscium | 4.13 | 13.71 | F7V | C |

**Appendix B: HabEx and LUVOIR-B Targets That Were Not Included in the ExEP List**

A comparison was made between the ExEP mission target list (Appendix A) and those for the HabEx and LUVOIR studies. For HabEx, we compare against the target lists from the HabEx Report (2020) Table D-1 ("*List of target stars obtained when optimizing the detection and spectral characterization of EECs with HabEx baseline architecture (4H), assuming 5 years of observations and no exozodi emission*") and Table D-2 ("*Illustrative list of target stars obtained when optimizing the detection and spectral characterization of EECs with HabEx baseline architecture (4H), randomly assigning individual stars exozodi levels and assuming 2 years of observations*"). The LUVOIR study report (LUVOIR Team, 2019) did not include a target list, however one was provided by Chris Stark (priv. Comm., 2019). Here we compare only against the list of LUVOIR-B (8-m) targets (omitting comparison to the 15-m LUVOIR-A) as its size was most similar to the Decadal-recommended aperture size of 6m.

In the following tables of stars, the main factor which led to exclusion from the ExEP list is indicated (the main reason for the low number of observable test cases for the fiducial exoplanets). The most common reasons are close binarity (<3″), low planet-star brightness ratios (typically for luminous stars) and low EEID compared to the minimum IWAs probed (typically for stars with low luminosities given their distances). In most cases where EEID is flagged, too few (<half) of the exoplanet cases (in planet radius, instellation, planet-star brightness ratio) were detectable for inclusion in the Tier A/B/C lists.

Component letters are given in order to clarify for which component the stellar parameters were evaluated - and pairs (e.g., AB, Aab) indicate that the calculations were done for the unresolved light of the pair. Stars that are spectroscopic binaries (SB) in the SB9 catalog (Pourbaix et al. 2004, most recent version on Vizier at https://cdsarc.cds.unistra.fr/viz-bin/cat/B/sb9) are indicated along with orbital periods. Separations in arcsecs are usually given from the Washington Double Star Catalog (WDS; Mason et al. 2001, most recent version on Vizier at https://cdsarc.cds.unistra.fr/viz-bin/cat/B/wds).

**LUVOIR-B:**

The following 44 stars were excluded from the ExEP list but appeared on the LUVOIR-B list. The primary reason for being excluded is mentioned, however most were excluded due either to close binarity, or the hypothetical HZ planets predominantly appearing at too small separations (and then the EEID angular separation at quadrature is listed).

HIP 10138 A   (binary: 2.6″)
HIP 10644 Aa  (binary: SB, P=10d)
HIP 12114 A   (binary: 1.7″)
HIP 17420     (EEID = 40 mas)
HIP 21770 A   (planet-star brightness ratio)
HIP 24186     (EEID = 32 mas)
HIP 25878     (planet-star brightness ratio)
HIP 28103     (planet-star brightness ratio)
HIP 34834     (planet-star brightness ratio)
HIP 36366 AB  (binary: 1.2″; planet-star brightness ratio)
HIP 37606 A   (binary: 0.1″; planet-star brightness ratio)
HIP 45333 Aab (binary: SB, P=16d; planet faintness)
HIP 46853 AB  (binary: 2.6″)
HIP 48331     (EEID = 36 mas)
HIP 58576 A   (binary: 1.4″)
HIP 64792 A   (binary: 2.5″)
HIP 65721     (binary: 2.9″)
HIP 67155     (EEID = 35 mas)
HIP 67927     (binary: SB, P=494d)
HIP 69972     (EEID = 49 mas)
HIP 70890     (EEID = 30 mas)
HIP 72848 AB  (binary: SB, P=125d)
HIP 76074     (EEID = 28 mas)
HIP 76829     (binary: 2.7″)
HIP 80686 Aab (binary: SB, P=13d)
HIP 80824     (EEID = 25 mas)
HIP 82860     (binary: SB, P=52d)
HIP 85295     (EEID = 43 mas)
HIP 85523     (EEID = 28 mas)
HIP 86162     (EEID = 32 mas)
HIP 86400 Aab (EEID = 52 mas)
HIP 88745 A   (binary: 0.2″,1.4″)

HIP 91768    (EEID = 35 mas)
HIP 94761 A   (EEID = 30 mas)
HIP 97944 Aa  (binary: SB, P=47d; EEID = 42 mas)
HIP 98036 A   (planet-star brightness ratio)
HIP 99701    (EEID = 40 mas)
HIP 101997  (EEID = 51 mas)
HIP 103096  (EEID = 32 mas)
HIP 106440  (EEID = 34 mas)
HIP 107556  (binary: SB, P=1d; planet-star brightness ratio)
HIP 109176 Aa (binary: SB, P=10d)
HIP 113576  (EEID = 39 mas)
HIP 117473  (EEID = 28 mas)

**HabEx:**

The following 31 stars were excluded from the ExEP list but appeared on the HabEx target list in the HabEx Study Report (2020). Note that most were excluded due to either close binarity (<3″) or low planet-star brightness ratios for the fiducial exoplanet cases tested.

HIP 1475 A   (EEID = 41 mas)
HIP 10644 Aa  (binary: SB, P=10d)
HIP 12114 A   (binary: 1.7″)
HIP 25878    (planet-star brightness ratio)
HIP 27321    (planet-star brightness ratio, dust; Beta Pic)
HIP 28103    (planet-star brightness ratio)
HIP 34834    (planet-star brightness ratio)
HIP 37279 A   (planet-star brightness ratio; binary: 3.8″, Procyon B)
HIP 39903    (binary: SB, P=899d)
HIP 45333 Aab (binary: SB, P=16d; planet faintness)
HIP 46509 A   (triple: SB/P=7.7yr, 67″)
HIP 57632    (binary: 1.4″)
HIP 58576 A   (binary: 1.4″)
HIP 64792 A   (binary: 2.5″)
HIP 65721    (binary: 2.9″)
HIP 67153    (binary: SB, P=10d; planet-star brightness ratio)
HIP 67927    (binary: SB, P=494d)
HIP 72848 AB  (binary: SB, P=125d)
HIP 76829    (binary: 2.7″)
HIP 80686 Aab (binary: SB, P=13d)

HIP 82860    (binary: SB, P=52d)
HIP 86400 Aab (EEID = 52 mas)
HIP 88745 A   (binary: 0.2″,1.4″)
HIP 92043    (planet-star brightness ratio)
HIP 95501 Aab (binary: P=3yr, 0.05″; planet-star brightness ratio)
HIP 97649    (planet-star brightness ratio)
HIP 98036 A   (planet-star brightness ratio)
HIP 102422   (planet-star brightness ratio)
HIP 107556   (binary: SB, P=1d; planet-star brightness ratio)
HIP 109176 Aa (binary: SB, P=10d)
HIP 113368   (planet-star brightness ratio)

**Appendix C: ExEP Target Stars That Were Not Included in Either the HabEx or LUVOIR-B Lists**

HIP 950 = Theta Sculptoris (F5V, d=21.7 pc; Tier C)
HIP 5896 A = Kappa Tucanae A (F5V, d=23.3 pc, 4.6″ binary; Tier C)
HIP 7751 B = p Eridani B (K2V, d=8.2 pc, 11.3″ binary; Tier A)
HIP 7751 A = p Eridani A (K2V, d=8.2 pc, 11.3″ binary; Tier A)
HIP 14879 A = Dalim (F6V, d=14.0 pc, 5.4″ binary; Tier C)
HIP 23835 = 104 Tauri (G1V, d=15.9 pc; spurious binary, see [WDS notes](); Tier C)
HIP 27072 B = Gamma Leporis B (K2.5V, d=8.9 pc; Tier B)
HIP 29650 = 71 Orionis (F5.5IV-V, d=21.8 pc, 8.0″ binary; Tier B)
HIP 34065 = HR 2667 (G1.5V, d=17.1 pc, 21.0″ binary; Tier B)
HIP 36439 = 22 Lyncis (F6V, d=20.4 pc; Tier B)
HIP 38423 A = 212 Puppis A (F5V, d=18.3 pc, 3.9″ binary; Tier C)
HIP 44897 = HR 3625 (G0IV-V, d=18.9 pc; Tier C)
HIP 45038 A = 13 Ursae Majoris A (F7V, d=20.5 pc, 4.3″ binary; Tier C)
HIP 50564 = 40 Leonis (F6IV-V, d=21.2 pc; Tier C)
HIP 51523 = HR 4134 (F9V, d=22.1 pc; Tier C)
HIP 64583 A = GJ 503A (F6V, d=18.2 pc, 4.6″ binary; Tier C)
HIP 69965 A = GJ 9476A (F9V, d=18.0 pc, 3.6″ binary; Tier C)
HIP 71681 = Toliman (K1V, d=1.3 pc, 5.3″ binary; Tier B)
HIP 72659 B = Xi Bootis B (K5V, d=6.7 pc, 5.2″ binary; Tier C)
HIP 77052 A = Psi Serpentis A (G5V, d=14.8 pc, triple - nearest 4.6″ away; Tier B)
HIP 84405 A = Guniibuu (K1V, d=6.0 pc, triple - nearest 5.1″ away; Tier B)
HIP 84405 B = Guniibuu B (K1V, d=5.9 pc, triple - nearest 5.1″ away; Tier B)
HIP 84720 A = 41 Arae A (G9V, d=8.8 pc, 10.6″ binary; Tier B)
HIP 84893 A = Aggia (F2V, d=17.5 pc, 4.1″ binary; Tier C)

HIP 88601 B = 70 Ophiuchi B (K4V, d=5.1 pc, 6.6″ binary; Tier B)
HIP 89348 = 36 Draconis (F5V, d=23.2 pc, 3.4″ binary; Tier C)
HIP 98959 = GJ 776 (G2V, d=17.9 pc; Tier C)
HIP 99461 A = GJ 783A (K2.5V, d=6.0 pc, 4.3″ comp., Tier C)
HIP 103389 = HR 8013 (F6V, d=21.1 pc, cold dusty disk; Tier C)
HIP 107350 = HN Pegasi (G0IV-V, d=18.1 pc, 3.0″ vis. comp., dist. 43″ comp., cold disk; Tier C)
HIP 114924 = HR 8853 (F8V, d=20.6 pc; Tier C)
HIP 114948 = HR 8843 (F6V, d=20.4 pc, cold dusty disk; Tier C)

**Appendix D: Problematic Targets like Binaries Requiring Further Analysis**

HIP 5896 A = Kappa Tuc A (4.6″, dm=2.66; Tier C)
HIP 14879 A = Dalim (5.4″, dm=3.21; Tier C)
HIP 29271 A = Alpha Men A (3.3″, dm=5.01; Tier C)
HIP 29650 = 71 Ori (8.0″, dm=6.00; Tier B)
HIP 38423 A = 212 Pup A (3.9″, dm=3.47; Tier C)
HIP 40693 = HD 69830 (6.9″, dm=13.44; Tier C)
HIP 45038 A = 13 UMa A (4.47″, dm=3.98; Tier C)
HIP 47080 A = 11 LMi (7.13″, dm=7.70; Tier B)
HIP 59199 A = Alchiba (3.07″, dm=7.75 [GaiaDR3]; Tier C)
HIP 64583 A = GJ 503A (4.6″, dm=5.28; Tier C)
HIP 64797 A = GJ 505A (7.6″, dm=2.84; Tier C)
HIP 69965 A = GJ 9476A (3.6″, dm=7.57; Tier C)
HIP 71681 = Toliman = Alpha Cen B (5.3″, dm=1.34; Tier B)
HIP 71683 = Rigil Kentaurus = Alpha Cen A (5.3″, dm=1.34; Tier B)
HIP 72659 A = Xi Boo A (5.2″, dm=2.19; Tier B)
HIP 72659 B = Xi Boo B (5.2″, dm=-2.19; Tier C)
HIP 77052 A = Psi Ser A (4.6″, dm=6.05; Tier B)
HIP 84405 A = Guniibuu = 36 Oph A (5.1″, dm=0.04; Tier B)
HIP 84405 B = Guniibuu B = 36 Oph B (5.1″, dm=-0.04; Tier B)
HIP 84720 A = 41 Ara A (SB?, 10.6″, dm=3.27; Tier B)
HIP 84893 A = Aggia = Xi Oph A (4.1″, dm=4.50; Tier C)
HIP 88601 A = 70 Oph A (6.6″, dm=1.95; Tier B)
HIP 88601 B = 70 Oph B (6.6″, dm=-1.95; Tier B)
HIP 89348 = 36 Dra (3.4″, dm=6.03; Tier C)
HIP 95447 = 31 Aql (4.2″, dm=6.47; Tier C)
HIP 99461 A = GJ 783A (4.3″, dm=6.19; Tier C)
HIP 107350 = HN Peg (3.0″, dm=16.10, interloper?, cold dust disk; Tier C)
HIP 111449 A = Upsilon Aqr A (6.1″, dm=5.20; Tier C)

**Appendix E: Stars Previously Reported to be Binary but Likely Spurious**

The following list of target stars had some indication of binarity previously reported in the literature (usually as either in the Washington Double Star Catalog or as a spectroscopic binary). However, upon closer examination of the literature either by ourselves or subsequent authors, it was concluded that either the initial observations were spurious, or there were other indicators suggestive that the star was likely single.

HIP 3765 = HR 222 (WDS, 2.7″ no dm; WDS J00484+0517 = HEI 202; spurious?; [WDS notes](); Tier A)

HIP 4151 = HR 244 (spurious SB1 from Abt & Levy 1976; multiple subsequent studies show RV to approximately constant; [SIMBAD notes](); Tier C)

HIP 15371 = Zeta2 Ret (WDS,0.0″ no dm; WDS J03182-6230 = BNU 2; spurious?; [WDS notes](); Tier A)

HIP 23835 = 104 Tau (WDS, 0.1″, dm=0.0; WDS J05074+1839 = A 3010; spurious?; [WDS notes](); Tier C)

HIP 61317 = Beta CVn = Chara (WDS, 0.1″, no dm; WDS J12337+4121 = BNU 4; spurious?; [WDS notes](); Tier A)

HIP 86736 = 58 Oph (WDS, 0.0″, dm=1.8; WDS J17434-2141 = OCC 401; spurious?; no WDS notes; [SIMBAD notes](); Tier C)

HIP 114046 = Lacaille 9352 (WDS, 0.1″, no dm; WDS J23059-3551 = WDK 4; spurious?, RV-detected multi-exoplanet system - no stellar companion; [Jeffers+2020](); Tier B)

## Appendix F: Table Column Descriptors and Notes

| tic_id | TIC designation in TESS Input Catalog (Stassun et al., 2019; 2019AJ....158..138S) |
|---|---|
| hip_name | HIP designation in Hipparcos Catalog (ESA 1997; 1997ESASP1200.....E) |
| hip_comname | Component letters provided for when multiple stars are associated with the HIP entry (following CCDM, WDS, or SIMBAD). |
| hd_name | HD designation in Henry Draper Catalog (Cannon & Pickering 1993; 1993yCat.3135....0C) |
| hr_name | HR designation in Bright Star Catalog, 5th Ed. (Hoffleit & Jaschek 1991; 1991bsc..book.....H) |
| gj_name | GJ designation in Preliminary Version of the Third Catalogue of Nearby Stars (CNS3; Gliese & Jahreiss 1991; 1991adc..rept.....G, Vizier catalog V/70A) or Fifth Catalogue of Nearby Stars (CNS5; Golovin et al. 2022; arXiv:2211.01449, https://dc.g-vo.org/CNS5) |
| constellation | Constellation ID using Greek letters or numbers following Bright Star Catalog, 5th Ed. (Hoffleit & Jaschek 1991; 1991bsc..book.....H) or Variable star ID from General Catalogue of Variable Stars Version GCVS 5.1 (Samus et al., 2007; 2017ARep...61...80S). Greek letter abbreviations follow SIMBAD and 3-letter constellation abbreviations follow IAU. |
| hostname | Common star ID, usually either (in order) IAU proper name, constellation ID (Greek letter, number, or variable star) written out in genitive form with Greek letter written in Latin alphabet, or ID from HD, HR, or GJ catalogs. In some cases, discovery ID is used for notable stars (e.g., Lacaille 8760). |
| sy_dist | Distance [parsecs], calculated as 1/parallax |
| sy_plx | Parallax [milliarcseconds] |
| sy_plxerr | Uncertainty in parallax [milliarcseconds] |

| sy_plx_reflink | Bibcode for reference for parallax |
|---|---|
| ra | Right Ascension (ICRS, epoch J2000) [deg] |
| dec | Declination (ICRS, epoch J2000) [deg] |
| sy_vmag | Apparent V(Johnson) magnitude |
| sy_vmagerr | Uncertainty in apparent V(Johnson) magnitude |
| sy_vmag_reflink | Bibcode for reference for apparent V(Johnson) magnitude |
| sy_bvmag | B-V (Johnson) color index (in magnitudes) |
| sy_bvmagerr | Uncertainty in B-V (Johnson) color index (in magnitudes) |
| sy_bvmag_reflink | Bibcode for reference for B-V (Johnson) color index |
| sy_rcmag | Apparent R(Cousins) magnitude |
| sy_rcmag_reflink | Bibcode for reference for apparent R(Cousins) magnitude |
| st_spectype | Spectral type |
| st_spectype_reflink | Bibcode for reference for spectral type |
| st_teff | Effective temperature [K] |
| st_tefferr | Uncertainty in effective temperature [K] |
| st_teff_reflink | Bibcode for reference for effective temperature |
| st_lum | $\text{Log}_{10}(L\_bol/L\_sun)$ of stellar bolometric luminosity (L_bol) normalized to Sun (L_sun) [dex] |
| st_lumerr | Uncertainty in $\text{Log}_{10}(L\_bol/L\_sun)$ [dex] |
| st_lum_reflink | Bibcode for reference for stellar bolometric luminosity |
| st_rad | Stellar radius in units of IAU nominal solar radius [Rsun] where 1 Rsun = 695700 km |
| st_diam | Stellar angular diameter [milliarcseconds] |
| st_mass | Stellar mass in units of solar mass [Msun] (1 Msun = nominal solar mass parameter (IAU 2015) / Newtonian constant of gravitation (CODATA 2018) = (GMsun)/G = 1.3271244e20 m^3 s^-2 / 6.6743e-11 m^3 kg^-1 s^-2 $\simeq$ |
|   | 1.9884e30 kg) |
| st_met | Stellar metallicity ($\log_{10}$ of ratio of iron or average of metals to hydrogen, normalized to solar; generally either [Fe/H] or [M/H]) [dex] |
| st_meterr | Uncertainty in stellar metallicity [dex] |
| st_metratio | Type of stellar metallicity (generally either [Fe/H] or [M/H]) |
| st_met_reflink | Bibcode for reference for stellar metallicity |
| st_logg | Log10 of stellar surface gravity in units of $\log_{10}$ [cm s^-2] |
| st_loggerr | Uncertainty in log(g) stellar surface gravity [dex] |
| st_logg_reflink | Bibcode for reference for stellar surface gravity |
| st_log_rhk | Stellar Ca II H & K chromospheric activity index $\log_{10}(R'\_HK)$ [dex] |
| st_log_rhk_reflink | Bibcode for reference for stellar Ca II H & K chromospheric activity index |
| st_eei_orbsep | Calculated Earth equivalent installation distance (EEID) [au] (calculated as square root of the stellar bolometric luminosity normalized to the Sun's) |
| st_eei_angsep | Calculated Angular Earth equivalent installation distance (EEID) [mas] (calculated as EEID in au divided by distance in parsecs) |
| st_etwin_bratio | Calculated planet-star brightness ratio for 1 R_Earth planet at EEID with assumed geometric albedo 0.2 at phase angle 90 degrees |
| st_etwin_rcmag | Calculated apparent R (Cousins) magnitude for 1 R_Earth planet at EEID with assumed geometric albedo 0.2 at phase angle 90 degrees (calculated as Rc(planet) = Rc(star) - $2.5\log_{10}$(planet-star brightness ratio)) |
| st_eei_orbper | Calculated orbital period for planet at EEID [days] |
| st_etwin_rvamp | Calculated radial velocity amplitude of star induced by 1 Mearth planet at EEID [cm/s] |
| st_etwin_astamp | Calculated astrometric amplitude of star induced by 1 Mearth planet at EEID [microarcseconds] |

| wds_designations | WDS designation for multiple system in Washington Double Star catalog (Mason et al. 2021; 2001AJ....122.3466M) |
|---|---|
| wds_comp | Component for star in WDS multiple system |
| wds_sep | Angular separation between star and other WDS component [arcseconds] |
| wds_deltamag | Magnitude difference between target star and other WDS component star [magnitudes] |
| planetsflag | Flag if star has one or more confirmed exoplanets in NASA Exoplanet Archive (Y=yes, N=no) [automatically generated in NExScI table, missing in preliminary ExEP table] |
| sy_diskflag | Flag if star has an infrared excess indicative of a dust disk, as measured by IRAS, Spitzer, Herschel, WISE, or LBTI (Y=yes, N=no). |
| sy_diskflag_reflink | Bibcode for reference for dust disk |
| target_group | Tier group for target star (A, B, C) |

Front Cover Credit: NASA/JPL-Caltech.

© 2023. California Institute of Technology. Government sponsorship acknowledged.

This document has been reviewed and determined not to contain export controlled technical data: JPL CL#23-0611.